# Magnetically Induced Metallic Phase in Superconducting Tantalum Films

Yongguang Qin, Carlos L. Vicente, and Jongsoo Yoon
*Department of Physics, University of Virginia, Charlottesville, VA 22903, U.S.A.*

We have studied the electronic transport properties of homogeneously disordered superconducting tantalum thin films in magnetic fields. The films exhibit three distinct transport regimes in the zero temperature limit which we identify as superconducting, metallic, and insulating phases. The metallic phase is unexpected. The transport characteristics of this metallic phase are found to be similar to those of MoGe films and high mobility dilute two-dimensional electrons or holes confined in semiconductor interface or transistor geometry.

Conventional treatments of electronic transport predict that the state of two-dimensional (2D) films can either be superconducting or insulating at zero temperature (T = 0) [1-5]. The fermionic picture, where the conductivity is mostly determined by the weakly decaying fermionic excitations, predicts that in the presence of weak disorder suppression of superconductivity leads to a decrease in the superconducting transition temperature ($T_c$) and causes a direct superconductor-insulator transition (SIT) in the T = 0 limit [2, 3]. In the strong disorder regime, where quantum fluctuations have the dominant effect, a direct SIT is also expected by the bosonic picture when the superconductivity is suppressed by disorder or magnetic fields (B) [3-5]. In the so called "dirty boson" model [4, 5], the superconducting phase is considered to be a condensate of Cooper pairs with localized vortices, and the insulating phase corresponds to a Bose-glass state which is a condensate of vortices with localized Cooper pairs.

In recent years, the report of an apparently metallic phase in homogeneously disordered MoGe films in the presence of weak magnetic fields [6-8] has triggered an intense interest in the possibility of a metallic ground state in 2D superconducting systems [9-16]. The metallic behavior is characterized by a drop in resistance ($\rho$) followed by a saturation to a finite value as T $\rightarrow$ 0. The "metallic" resistance can be orders of magnitude smaller than the normal state resistance ($\rho_n$) implying that the metallic state exists as a separate phase rather than a point in the phase diagram. The nature of this metallic phase is not well understood. While some theories [6, 7] argue that the metallic behavior is due to a finite temperature effect and there should be no true metallic ground state at T = 0, other models [8-13] obtain a true metallic ground state when the superconductivity is suppressed. Proposed origins for the metallic phase include bosonic interactions in the non-superconducting phase [8, 9], contribution of fermionic quasiparticles to the conduction [10, 11], and quantum phase fluctuation [12, 13].

Further interest in the metallic ground state has arisen from the observation of similar metallic behavior in high mobility dilute 2D electrons or holes confined in semiconductor interface or transistor geometry such as GaAs/AlGaAs, Si/SiGe, and Si-MOSFET's (Metal-Oxide-Semiconductor-Field-Effect-Transistors) [17]. These experiments imply that the metallic behavior is common to high mobility dilute 2D electron systems where the Coulomb interaction energy, which is ignored in the traditional scaling theory of localization [1], is the dominant energy scale. On the other hand, among many homogeneously disordered 2D superconducting systems that have been investigated, MoGe has been the only material system where a metallic behavior has been reported. In this Letter, we report a metallic behavior observed in homogeneously disordered Ta thin films when the superconductivity is suppressed by magnetic fields. We found that the metallic regime can be identified by nonlinear voltage-current (VI) characteristics that are qualitatively different from the zero or low B superconducting and high B insulating regimes.

The Ta films are dc sputter deposited on Si substrates. The sputter chamber is baked at ~ 110 °C for 2 days reaching a base pressure of ~ $10^{-8}$ Torr. The chamber and Ta source were cleaned by pre-sputtering for ~ 30 minutes at a rate of ~ 1 nm/sec. Films were grown at a rate of ~ 0.05 nm/sec at an Ar pressure of ~ 4 mTorr. Using a rotatable substrate holder, up to 12 films each with a different thickness can be grown without breaking the vacuum. The films are patterned into a bridge, 1 mm wide and 5 mm long, for the standard four point measurement with a shadow mask. Sheet resistances at 4.2K ($\rho_{4.2K}$) for films from the same batch have shown a smooth and monotonic increase with decreasing film thickness. There were noticeable batch-to-batch variations in $\rho_{4.2K}$ for nominally the same thickness films. However, the transport characteristics of a film such as the $T_c$ and T-dependence of $\rho$, are uniquely determined by its sheet resistance. Special care was taken to avoid noise by using an in-line noise filter ($\pi$-filter) or performing the measurements inside an electromagnetically shielded room.

Because Ta is a high Z-number element, X-ray scattering is a powerful tool to probe the structure. The inset in Fig. 1 shows the X-ray diffraction patterns of 15, 10, 5 nm thick films, and the bare substrate. 15 and 10 nm thick films show broad and shallow peaks at ~ 33° and ~ 37°. They are due to local tetragonal (known as $\beta$-phase) and body-centered-cubic ($\alpha$-phase) correlation [18], respectively, over a length scale of order of 1 nm. These peaks are smaller by several orders of magnitude and ~ 10 times broader than the peaks of thick films in the bulk limit. Materials exhibiting such broad and shallow X-ray peaks are known to give concentric electron diffraction rings [19] that are often considered as an evidence of an amorphous structure of a thin film [20]. The trace for the 5 nm thick film does not show any sign of local atomic correlation indicating that the structure is highly amorphous. We focus on films with thickness less than ~ 5 nm.

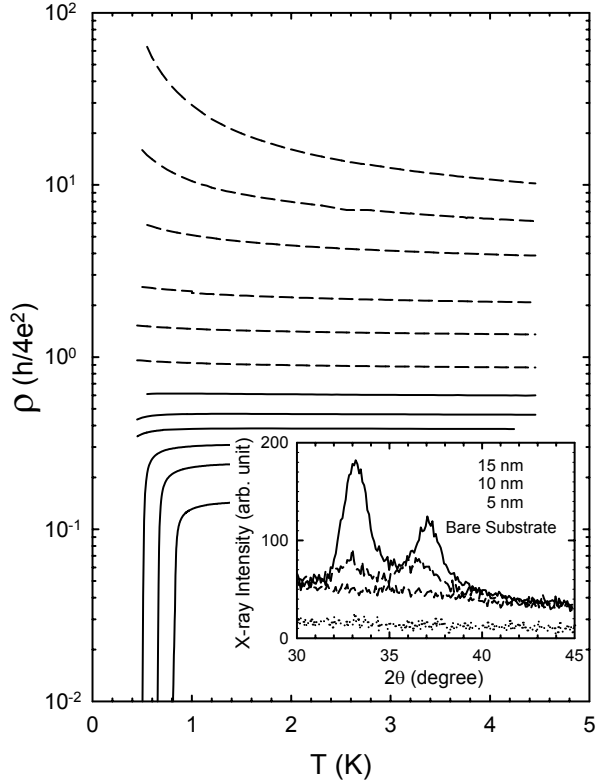

**FIG. 1**. The resistance is measured at B = 0 with a dc current in the range 1-10 nA that is within the linear response regime. The thicknesses of the 12 films are, from the top, 1.9, 2.0, 2.1, 2.3, 2.5, 2.8, 3.1, 3.4, 3.7, 4.0, 4.5, and 5.5 nm. The dashed lines are for the insulating phase and the solid lines are for the superconducting phase. Inset: X-ray diffraction patterns of 15, 10, 5 nm thick Ta films, and a bare Si substrate.

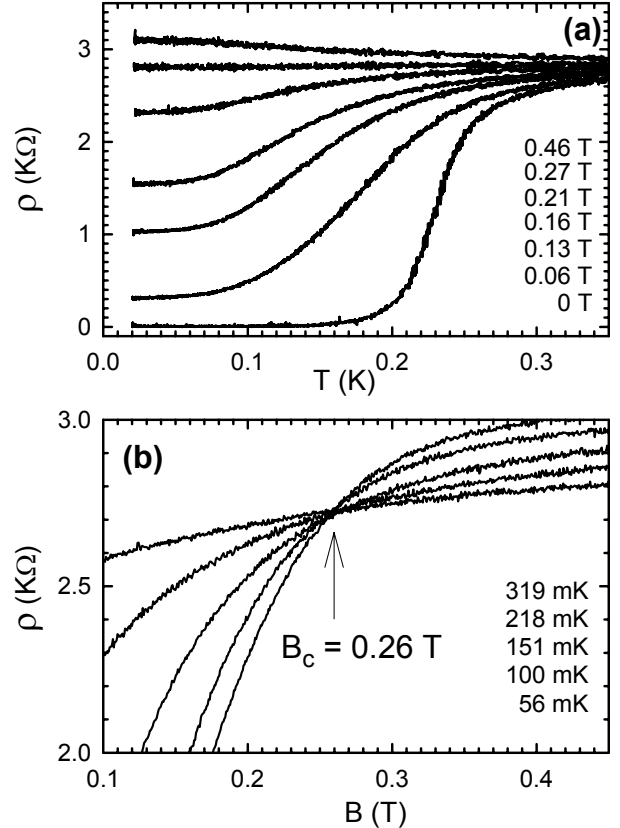

**FIG. 2. (a)** Sheet resistances are measured with a dc current of 1 nA at the indicated magnetic fields (from top to bottom). **(b)** Magnetoresistances are measured with a dc current of 1 nA at the indicated temperatures. At fields above (below) the crossing point, the top (bottom) trace corresponds to the lowest temperature. The crossing point at 0.26 T marked by an arrow defines $B_c$.

The T-dependence of ρ in the absence of magnetic field for 12 films in the thickness range 1.9 – 5.5 nm is shown in Fig. 1. For bottom traces (solid lines), ρ stay T-independent as T is decreased their $T_c$ are reached at which ρ drops sharply. For the top traces (dashed lines), on the other hand, the temperature coefficient of resistance (dρ/dT) is always negative even at 4 K, which is characteristic of an insulator. The behaviors shown in Fig. 1 represent a B = 0 thickness-tuned SIT that have been observed in many other homogeneously disordered films [21]. On the superconducting side, the $T_c$ decrease continuously towards 0 K with decreasing film thickness, and their superconducting transitions show no sign of reentrant behavior. These superconducting characteristics are typical of homogeneously disordered thin films, and consistent with the results of our X-ray measurements.

We have studied the effect of B, perpendicular to the sample plane, on transport for 5 different films in the 3.5 - 5.75 nm thickness range. Qualitatively similar results are obtained in all 5 samples, but the data presented in the rest of the Letter is from a 5 nm thick film. The correlation length of the film is estimated to be $\xi(T=0) = \sqrt{\Phi_o / 2\pi B_c} \approx 35$ nm, where $\Phi_o$ is the flux quantum, indicating that the film is in the 2D limit. We have used Bc=0.26 T, the field at which ρ is 90% of the normal state resistance; this field coincides with the field above which the sample turns insulating.

Figure 2(a) shows the effect of B. At B = 0 (the bottom trace), the film becomes superconducting with the mean field $T_c \approx 0.23$ K, and the resistance at our lowest temperature of 0.02 K is "zero" within the accuracy of the measurements. However, in the presence of a magnetic field, 0.06 T for example, the resistance decreases with decreasing T but saturates to a measurably large value. This resistance drop followed by saturation to a finite value as T → 0, is found to persist to a field up to B~ 0.27 T. In this field regime, based on the measured data down to ~ 0.02 K, the resistance at T = 0 extrapolates to a finite value implying that the ground state is metallic. The resistance in the low T limit increases towards $\rho_n$ with increasing B, and at B = 0.27 T it becomes almost T-independent over the entire T range. At higher fields (the top trace in Fig. 2(a)) $d\rho/dT$ becomes negative which we take as an indication for an insulator. We phenomenologically identify the three distinct transport regimes as "superconducting", "metallic", and "insulating". The transition from the metallic to insulating phase occurs at a well defined "critical" field $B_c$. This is shown in Fig. 2(b) where $\rho$ is plotted against B at five temperatures in the range 0.056 – 0.319 K. In this plot, $B_c$ appears as a single crossing point. This is the direct consequence of the fact that, in the covered T range, $\rho$ decreases with decreasing T for B < $B_c$, increases with decreasing T for B > $B_c$, and is T-independent at $B_c$.

Remarkably different nonlinear VI characteristics are found in the three phases. The VI curves in the superconducting phase, as shown in Fig. 3(a) for B = 0, are characterized by hysteretic voltage jumps, possibly due to pinning-depinning of vortices. With increasing I, the superconducting or "zero" resistance state is abruptly quenched at a well-defined critical current $I_{c1}$. With decreasing I, the voltage suddenly drops to "zero" at another critical current $I_{c2}$ which is clearly below $I_{c1}$. In the metallic phase, the VI's are continuous and reversible as shown in the inset of Fig. 3(a), and the differential resistance (dV/dI) increases with increasing I in either direction (the bottom trace in Fig. 3(b)). In the insulating phase (B > $B_c$, the top trace), dV/dI decreases with increasing I in either direction. Near the metal-insulator boundary (B ≈ $B_c$, the middle trace) dV/dI is constant implying a linear VI. It is interesting to point out that the characteristics of the metallic transport, both T-dependence of $\rho$ and nonlinear VI characteristics, are strikingly similar to those of the metallic behavior of a two dimensional hole system (2DHS) [22, 23]. Such close similarities are intriguing considering the fundamental difference between the two systems: one is bosonic, as manifested by the superconductivity at B = 0; the other is fermionic, as evidenced by the quantum Hall effect under perpendicular magnetic fields.

The qualitative differences in the nonlinear VI for the three phases provide a new classification criterion that can be more convenient than the sign of $d\rho/dT$, as it allows one to determine the phase from VI data taken at a single temperature. We note that for the metallic and insulating phases, the phase identification based on the nonlinear VI is always consistent with that based on the sign of $d\rho/dT$. Identifying the superconducting phase by a hysteretic VI is somewhat artificial. Strictly speaking, in 2D the true superconductivity is believed to exist only at T = 0, and assumed to be destroyed by any finite current [24].

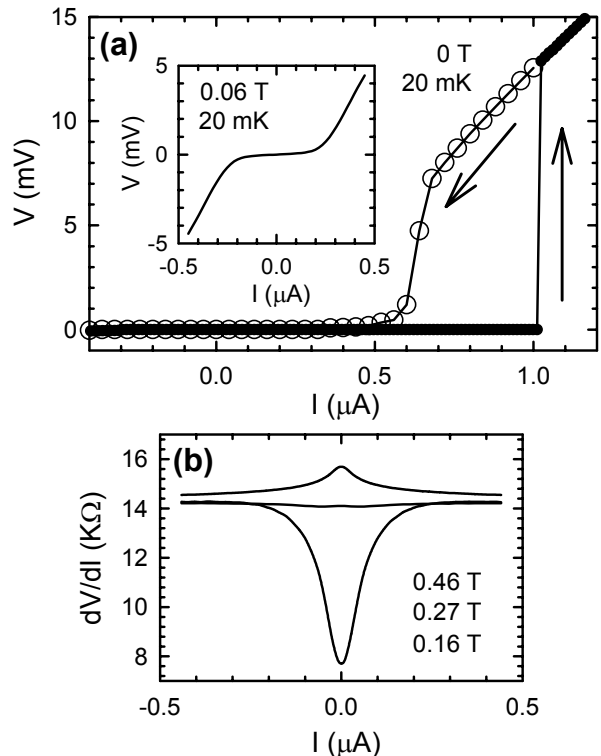

**FIG. 3. (a)** Hysteretic voltage jumps in the superconducting phase at B = 0 and T = 20 mK. Arrows are to indicate the current increasing or decreasing direction. Inset: Smooth and reversible VI curves in the metallic phase at B = 0.06 T and T = 20 mK. **(b)** Differential resistance across the metal-insulator boundary is shown. The magnetic field for each trace is indicated from top to bottom.

However, a hysteretic VI represents transport characteristics that are clearly distinguishable from the metallic phase where VI's are smooth and reversible. Moreover, when VI's are hysteretic, the resistance is always extremely small and the state can be considered superconducting the sense that it has an "immeasurably" small resistance. Identifying the superconducting phase by the hysteretic VI, the critical field $B_c^{SM}$ for superconductor-metal boundary is determined to be ~ 0.04 T. Both $B_c^{SM}$ and $B_c$ are found to increase with decreasing $\rho_n$, and for a film with $\rho_n \approx 1$ KΩ we found that $B_c^{SM} \approx 0.2$ T, and $B_c \approx 1.0$ T.

The origin of the nonlinear VI's in the metallic and insulating phase is not clear at present. In the insulating phase they might be caused by current-induced delocalization of localized Cooper pairs, and in the metallic phase by a similar effect of current on vortices. Alternatively, the nonlinear VI's could be caused by electron heating by the bias current: a temperature rise in the metallic phase would result in an increase in $\rho$, and in the insulating phase a decrease. We would like to point out that in the metallic regime of a 2DHS where the T-dependence of $\rho$ is nonmonotonic ($d\rho/dT > 0$ at low temperatures, and $d\rho/dT < 0$ at high temperatures), $dV/dI$ has been found to increase with increasing I at both low and high temperatures [23]. This demonstrates that the nonlinear VI of a 2DHS is not caused by electron heating, and suggests that the nonlinearity in the VI might be an intrinsic property of our system.

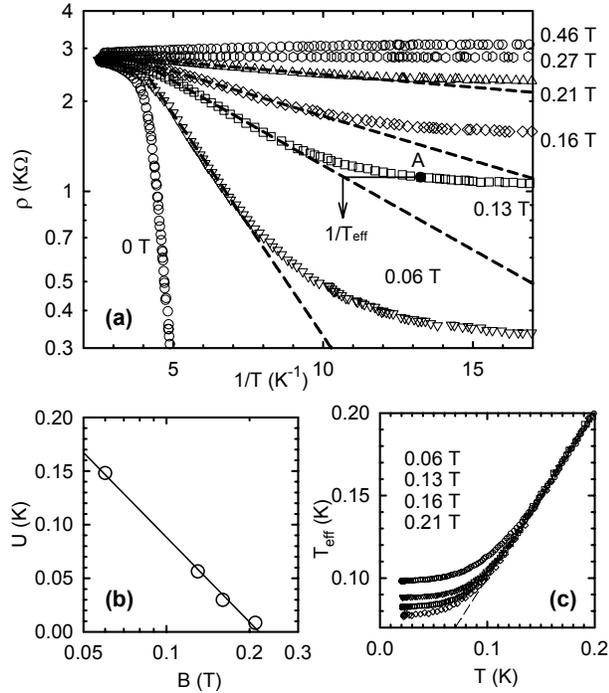

**FIG. 4. (a)** Arrhenius plot of the data shown in Fig. 2(a). Magnetic field for each trace is indicated. The high temperature data are fitted to the activated behavior, $\rho \sim e^{-U(B)/T}$ (dashed lines). $T_{eff}$ is determined as illustrated for a data point A. **(b)** Activation energy $U(B)$. Solid line is a fit to the form, $U(B) \sim \ln(B_o/B)$. **(c)** $T_{eff}$ is plotted against sample stage temperature. The magnetic fields indicated are for traces from top to bottom at low temperature end. The dashed line is to indicate the activated behavior.

Finally, we turn to the issue on the thermal decoupling effect of electrons in the metallic phase from the stage where the sample is thermally anchored. The effect, if strong enough, would make the sample resistance to appear to saturate as $T \rightarrow 0$. In MoGe films this issue has been addressed [16] by extracting the "effective" temperature ($T_{eff}$) at which the electrons would have to be if the data in the metallic phase were to follow the same activated behavior, $\rho \sim e^{-U(B)/T}$, that is observed at higher T. The activation energy $U(B)$ corresponds to the thermal energy to unbind pairs of dislocations and antidislocations of the vortex lattice [25]. Here we analyze our data in the same manner. The value of $U(B)$ obtained from a fit to the high T activated behavior (dashed lines in Fig. 4(a)) is plotted in Fig. 4(b). The B-dependence is in a reasonably good agreement with the expected form [25], $U(B) \sim \ln(B_o/B)$. From a fit (solid line) we obtain $B_o = 0.22$ T which is very close to $B_c = 0.26$ T as predicted. $T_{eff}$ is determined by horizontally projecting the measured data onto the dashed lines as illustrated in Fig. 4(a) for a data point A. We plot $T_{eff}$ against the stage temperature in Fig. 4(c). The plot shows two main results of this analysis. One is that the limiting value of $T_{eff}$ depends on B, and the other is that the deviation from the activated behavior occurs at lower T in the presence of higher B where the sample resistance, hence the power dissipated, and the heating in the sample are greatest. These results suggest that the apparent metallic behavior is likely intrinsic to the sample and not entirely the consequence of electrons not cooling.

In summary, we have reported that magnetic fields can drive Ta thin films from superconducting to metallic, and then to insulating phase. Each phase can be unambiguously identified by their nonlinear VI's. The unexpected metallic phase shows transport characteristics that are similar to the metallic behavior observed in a 2DHS in GaAs/AlGaAs.

Authors acknowledge fruitful discussions with V. Galitski, and D. Louca and K. Unruh on interpreting the X-ray data. This work is supported by NSF. A portion of this work was performed at the National High Magnetic Field Laboratory.